\documentclass{jfm}
\usepackage{graphicx}
\usepackage{epstopdf, epsfig}

\usepackage{amsmath}
\usepackage[colorlinks, citecolor=blue]{hyperref}

\usepackage{siunitx}

\usepackage{xcolor}

\usepackage[normalem]{ulem}
\usepackage[textwidth=\dimexpr\textwidth-2cm\relax]{todonotes}
\makeatletter
\@mparswitchfalse%
\makeatother
\normalmarginpar 

\shorttitle{The role of viscosity on drop impact forces on non-wetting surfaces}
\shortauthor{V. Sanjay, B. Zhang, C. Lv, and D. Lohse}

\title{The role of viscosity on drop impact forces on non-wetting surfaces}
\author{Vatsal Sanjay\aff{1}
	\corresp{\email{vatsalsanjay@gmail.com}},
	Bin Zhang\aff{2}
	\corresp{\email{binzhang0710@gmail.com}},
	Cunjing Lv\aff{2}
	\corresp{\email{cunjinglv@mail.tsinghua.edu.cn}},
	\and Detlef Lohse{\aff{1}$^{,}$\aff{3}}
	\corresp{\email{d.lohse@utwente.nl}}}

\affiliation{\aff{1}Physics of Fluids Group, Max Planck Center for Complex Fluid Dynamics, Department of Science and Technology, and J. M. Burgers Centre for Fluid Dynamics, University of Twente,  P. O. Box 217, 7500 AE Enschede, The Netherlands\aff{2}Department of Engineering Mechanics, AML, Tsinghua University, Beijing 100084, China\aff{3}Max Planck Institute for Dynamics and Self-Organization, Am Fassberg 17, 37077 G\"{o}ttingen, Germany}

\newcommand{\Wen}{\mathit{We}}

\begin{document}
	\maketitle
	
	\begin{abstract}
		
		A liquid drop impacting a rigid substrate undergoes deformation and spreading due to normal reaction forces, which are counteracted by surface tension. On a non-wetting substrate, the drop subsequently retracts and takes off.
		Our recent work (Zhang et al., \textit{Phys. Rev. Lett.}, vol. 129, 2022, 104501) revealed two peaks in the temporal evolution of the normal force $F(t)$ -- one at impact and another at jump-off. The second peak coincides with a Worthington jet formation, which vanishes at high viscosities due to increased viscous dissipation affecting flow focusing.
		In this article, using experiments, direct numerical simulations, and scaling arguments, we characterize both the peak amplitude $F_1$ at impact and the one at take off ($F_2$) and elucidate their dependency on the control parameters: the Weber number $We$ (dimensionless impact kinetic energy) and the Ohnesorge number $Oh$ (dimensionless viscosity). 
		The first peak amplitude $F_1$ and the time $t_1$ to reach it depend on inertial timescales for low viscosity liquids, remaining nearly constant for viscosities up to 100 times that of water. For high viscosity liquids, we balance the rate of change in kinetic energy with viscous dissipation to obtain new scaling laws: $F_1/F_\rho \sim \sqrt{Oh}$ and $t_1/\tau_\rho \sim 1/\sqrt{Oh}$, where $F_\rho$ and $\tau_\rho$ are the inertial force and time scales, respectively, which are consistent with our data.
		The time $t_2$ at which the amplitude $F_2$ appears is set by the inertio-capillary timescale $\tau_\gamma$, independent of both the viscosity and the impact velocity of the drop. However, these properties dictate the magnitude of this amplitude.

	\end{abstract}
	
	
	\section{Introduction} \label{sec:intro}
	
	\begin{figure}
		\centering
		\includegraphics[width=\textwidth]{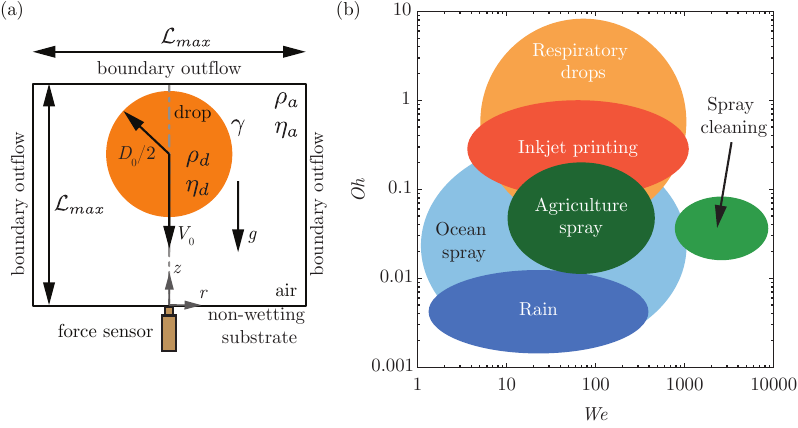}
		\caption{(a) Problem schematic with an axisymmetric computational domain used to study the impact of a drop with diameter $D_0$ and velocity $V_0$ on a non-wetting substrate. In the experiments, we use a quartz force sensor to measure the temporal variation of the impact force. The subscripts $d$ and $a$ denote the drop and air, respectively, to distinguish their material properties, which are the density $\rho$ and the dynamic viscosity $\eta$. The drop–air surface tension coefficient is $\gamma$. The grey dashed-dotted line represents the axis of symmetry, $r = 0$. Boundary air outflow is applied at the top and side boundaries (tangential stresses, normal velocity gradient, and ambient pressure are set to zero). The domain boundaries are far enough from the drop not to influence its impact process ($\mathcal{L}_\text{max} \gg D_0$, $\mathcal{L}_\text{max} = 8R$ in the worst case). (b) The phase space with control parameters: the Weber number ($We$: dimensionless kinetic energy) and the Ohnesorge number ($Oh$: dimensionless viscosity), exemplifying different applications.}
		\label{fig:schematic}
	\end{figure}
	\begin{figure}
		\centering
		\includegraphics[width=\textwidth]{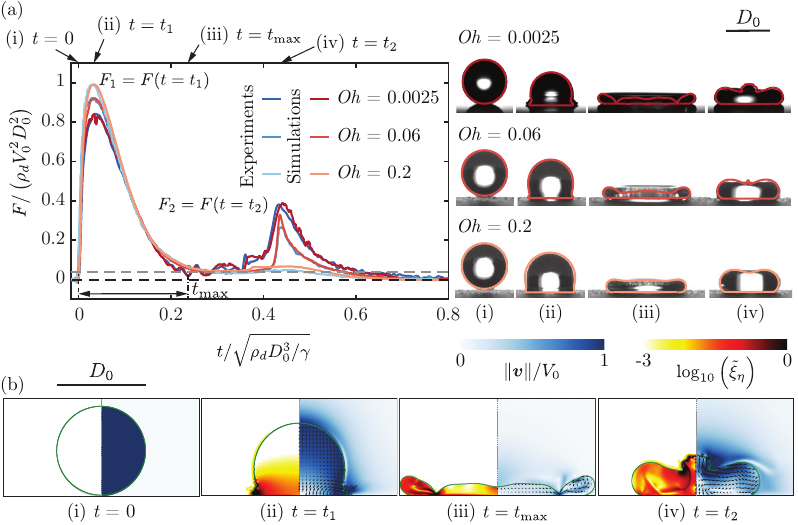}
		\caption{Comparison of the drop impact force $F(t)$ obtained from experiments and simulations for the three typical cases with impact velocity $V_0 = 1.2\,\si{\meter}/\si{\second}, 0.97\,\si{\meter}/\si{\second}, 0.96\,\si{\meter}/\si{\second}$, diameter $D_0 = 2.05\,\si{\milli\meter}, 2.52\,\si{\milli\meter}, 2.54\,\si{\milli\meter}$, surface tension $\gamma = 72\,\si{\milli\newton}/\si{\meter}, 61\,\si{\milli\newton}/\si{\meter}, 61\,\si{\milli\newton}/\si{\meter}$ and viscosity $\eta_d = 1\,\si{\milli\pascal\second}, 25.3\,\si{\milli\pascal\second}, 80.2\,\si{\milli\pascal\second}$. These parameter give $Oh = 0.0025, 0.06, 0.2$ and $We = 40$.
			For the three cases, the two peak amplitudes, $F_1/(\rho_dV_0^2D_0^2) \approx$ 0.82, 0.92, 0.99 at $t_1 \approx 0.03\sqrt{\rho_dD_0^3/\gamma}$ and $F_2/(\rho_dV_0^2D_0^2) \approx$ 0.37, 0.337, 0.1 at $t_2 \approx 0.42\sqrt{\rho_dD_0^3/\gamma}$, characterize the inertial shock from impact and the Worthington jet before takeoff, respectively. 
			The drop reaches the maximum spreading at $t_{\text{max}}$ when it momentarily stops and retracts until $0.8\sqrt{\rho_dD_0^3/\gamma}$ when the drop takes off ($F = 0$). The black and gray dashed lines in panel (a) mark $F = 0$ and the resolution $F = 0.5\,\si{\milli\newton}$ of our piezoelectric force transducer, respectively.
			(b) Four instances are further elaborated through numerical simulations for ($We = 40, Oh = 0.0025$), namely (i) $t = 0$ (touch-down), (ii) $t = 0.03\sqrt{\rho_dD_0^3/\gamma}$ ($t_1$), (iii) $t = 0.2\sqrt{\rho_dD_0^3/\gamma}$ ($t_{\text{max}}$), and (iv) $t = 0.42\sqrt{\rho_dD_0^3/\gamma}$ ($t_2$).
			The insets of panel (a) exemplify these four instances for the three representative cases illustrated here. The experimental snapshots are overlaid with the drop boundaries from simulations. 
			We stress the excellent agreement between experiments and simulations without any free parameters.
			The left part of each numerical snapshot shows (on a $\log_{10}$ scale) the dimensionless local viscous dissipation function $\tilde{\xi}_\eta \equiv \xi_\eta D_0/\left(\rho_dV_0^3\right) = 2Oh\left(\boldsymbol{\tilde{\mathcal{D}}:\tilde{\mathcal{D}}}\right)$, where $\boldsymbol{\mathcal{D}}$ is the symmetric part of the velocity gradient tensor, and the right part the velocity field magnitude normalized with the impact velocity. The black velocity vectors are plotted in the center of mass reference frame of the drop to clearly elucidate the internal flow.  Also see supplementary videos SM1-SM3.}
		\label{fig:summary}
	\end{figure}
	
	Drop impacts have piqued the interest of scientists and artists alike for centuries, with the phenomenon being sketched by \citet{da1508notebooks} in the early 16$^{\text{th}}$ and photographed by \citet{worthington1876, worthington1876b} in the late 19$^{\text{th}}$ century. It is, indeed, captivating to observe raindrops hitting a solid surface \citep{kim2020raindrop, lohse-2020-pnas} or ocean spray affecting maritime structures \citep{berny2021statistics, villermaux2022bubbles}. The phenomenology of drop impact is extremely rich, encompassing behaviors such as drop deformation \citep{Biance2006, molavcek2012quasi, chevy2012liquid}, spreading \citep{laan2014maximum, Wildeman2016}, splashing \citep{xu2005drop, riboux2014experiments, thoraval2021nanoscopic}, fragmentation \citep{villermaux2011drop, villermaux2020fragmentation}, bouncing \citep{Richard2000, kolinski-2014-epl, Jha2020, chubynsky-2020-prl, sharma-2021-jfm, sanjay_chantelot_lohse_2023}, and wetting \citep{degennes-1985-rmp, fukai-1995-pof, quere-2008-arms, Bonn2009}. These behaviors are influenced by the interplay of inertial, capillary, and viscous forces, as well as additional factors like non-Newtonian properties \citep{bartolo2005retraction, bartolo2007dynamics,  smith-2010-prl, gorin-2022-langmuir} of the liquid and even ambient air pressure \citep{xu2005drop}, making the parameter space for this phenomenon both extensive and high-dimensional. 
	
	Naturally, even the process of a Newtonian liquid drop impacting a rigid substrate is governed by a plethora of control parameters, including but not limited to the drop's density $\rho_d$, diameter $D_0$, velocity $V_0$, dynamic viscosity $\eta_d$, surface tension $\gamma$, and acceleration due to gravity $g$ (figure~\ref{fig:schematic}a). 
	To navigate this rich landscape, we focus on two main dimensionless numbers that serve as control parameters (figure~\ref{fig:schematic}b): the Weber number $We$, which is the ratio of inertial to capillary forces and is given by
	
	\begin{align}
		\label{Eq:WeDefinition}
		We = \frac{\rho_d V_0^2 D_0}{\gamma},
	\end{align}
	
	\noindent and the Ohnesorge number $Oh$, which captures the interplay between viscous damping and capillary oscillations, offering insights into how viscosity affects the drop's behavior upon impact,
	
	\begin{align}
		\label{Eq:OhDefinition}
		Oh = \frac{\eta_d}{\sqrt{\rho_d \gamma D_0}}.
	\end{align}
	
	\noindent Additionally, the Bond number
	
	\begin{align}
		\label{Eq:BoDefinition}
		Bo = \frac{\rho_dgD_0^2}{\gamma}
	\end{align}
	
	\noindent compares gravity to inertial forces and is needed to uniquely define the non-dimensional problem.  
	
	The drop impact is not only interesting from the point of view of fundamental research but also finds relevance in inkjet printing \citep{lohse2022fundamental}, the spread of respiratory drops carrying airborne microbes \citep{bourouiba2021fluid, ji2021compound, pohlker2023respiratory}, cooling applications \citep{kim2007spray, shiri2017heat, jowkar2019rebounding}, agriculture \citep{bergeron2000controlling, bartolo2007dynamics, kooij2018determines, sijs2020effect, he2021optimization, hoffman2021controlling}, criminal forensics \citep{smith2018influence, smith-2018-curropincolloidinterfacesci}, and many other industrial and natural processes \citep{rein-1993-fluiddynres, yarin2006drop, Tuteja2007, Cho2016, Josserand2016, Yarin2017, Liu2017, Hao2016, Yarin2017, Wu2020}. For these applications, it is pertinent to understand the forces involved in drop impacts, as these forces can lead to soil erosion \citep{Nearing1986} or damage to engineered surfaces \citep{Ahmad2013, Amirzadeh2017, Gohardani2011}. We refer the readers to \citet{cheng2021drop} for an overview of the recent studies unraveling drop impact forces; see also \citet{Li2014, Soto2014, Philippi2016, Zhang2017, Gordillo2018, Mitchell2019, Zhang2019}. 
	
	These forces have been studied by \citet{zhang2022impact}, employing experiments and simulations and deriving scaling laws. 
	A liquid drop impacting a non-wetting substrate undergoes a series of phases--spreading, recoiling, and potentially rebounding \citep{chantelot2018rebonds}--driven by the normal reaction force exerted by the substrate (figure~\ref{fig:summary}). 
	The moment of touch-down (see figures~\ref{fig:summary}a,b $t = 0$ to $t_1$) \citep{wagner1932stoss, Philippi2016, Gordillo2018} is not surprisingly associated with a pronounced peak in the temporal evolution of the drop impact force $F(t)$ owing to the sudden deceleration as high as $100$ times the acceleration due to gravity \citep{Clanet2004} (figure~\ref{fig:summary}a, $F_1/(\rho_dV_0^2D_0^2) \approx$ 0.82, 0.92, 0.99 for $Oh = 0.0025, 0.06,\,\text{and}\,0.2$, respectively; at $t_1 \approx 0.03\sqrt{\rho_dD_0^3/\gamma}$.
	The force diminishes as the drop reaches its maximum spreading diameter (figure~\ref{fig:summary}a,b $t = t_m$). 
	\citet{zhang2022impact} revealed that also the jump-off is accompanied by a peak in the normal reaction force, which was up to then unknown (figure~\ref{fig:summary}a, 
	$F_2/(\rho_dV_0^2D_0^2) \approx$ 0.37, 0.337, 0.1 for $Oh = 0.0025, 0.06,\,\text{and}\,0.2$, respectively; for the second force peak amplitude--at time $t_2 \approx 0.42\sqrt{\rho_dD_0^3/\gamma}$ after impact)
	The second peak in the force also coincides with the formation of a Worthington jet, a narrow upward jet of liquid that can form due to flow focusing by the retracting drop (figure~\ref{fig:summary}a,b $t = t_2$). Under certain conditions ($We \approx 9$, $Oh = 0.0025$), this peak can be even more pronounced than the first. This discovery is critical for superhydrophobicity which is volatile and can fail due to external disturbances such as pressure \citep{Lafuma2003, Callies2005, Sbragaglia2007, Li2017}, evaporation \citep{Tsai2010, Chen2012, Papadopoulos2013},  mechanical vibration \citep{Bormashenko2007}, or the impact forces of prior droplets \citep{Bartolo2006Bouncing}.   
	
	In contrast to our prior study \citet{zhang2022impact}, which fixed the Ohnesorge number to that of a 2 mm diameter water drop (Oh = 0.0025), our present investigation reported in this paper explores a broader parameter space. We systematically and independently vary the Weber and Ohnesorge numbers, extending the range of $Oh$ to as high as 100. This comprehensive approach enables us to develop new scaling laws and provides a more unified understanding of the forces involved in drop impact problems. Our findings are particularly relevant for applications with varying viscosities and impact velocities (figure~\ref{fig:schematic}). 
	
	The structure of this paper is as follows: \S\ref{sec:Methods} briefly describes the experimental and numerical methods. \S\ref{sec:FirstPeak} and  \S\ref{sec:SecondPeak} offer detailed analyses of the first and second peaks, respectively, focusing on their relationships with the Weber number ($We$) and the Ohnesorge number ($Oh$). Conclusions and perspectives for future research are presented in Section~\ref{sec:Conclusion}.
	
	\section{Methods}\label{sec:Methods}
	
	\subsection{Experimental method}
	
	\begin{table}
		\begin{center}
			\def~{\hphantom{0}}
			\begin{tabular}{lccc}
				glycerol & $\rho_d$ & $\eta_d$  & $\gamma$\\
				(wt \%) &(kg/m$^{3}$) &(mPa.s)& (mN/m)\\[3pt]
				0 & 1000 & 1 & 72 \\
				50 & 1124 & 5 & 61 \\
				63 & 1158 & 10 & 61 \\
				74 & 1188 & 25.3 & 61 \\
				80 & 1200 & 45.4 & 61 \\
				85 & 1220 & 80.2 & 61 \\
			\end{tabular}
			\caption{Properties of the water-glycerol mixtures used in the experiments. $\rho_d$ and $\eta_d$ are the density and viscosity of the drop, respectively and $\gamma$ denotes the liquid-air surface tension coefficient. These properties are calculated using the protocol provided in \citet{cheng2008formula, volk2018density}.}
			\label{tab:table00}
		\end{center}
	\end{table}
	
	In the experimental setup, shown schematically in figure~\ref{fig:schematic}(a), a liquid drop impacts a superhydrophobic substrate. For water drops, such a surface is coated with silanized silica nanobeads with a diameter of $20\,\si{\nano\meter}$ (Glaco Mirror Coat Zero; Soft99) resulting in the advancing and receding contact angles of $167 \pm 2^{\circ}$ and $154 \pm 2^{\circ}$, respectively \citep{Gauthier2015, Li2017}. On the other hand, for viscous aqueous glycerin drops, the upper surface is coated with an acetone solution of hydrophobic beads (Ultra ever Dry, Ultratech International, a typical bead size of 20 nm), resulting in the advancing and receding contact angles of $166 \pm 4^{\circ}$ and $159 \pm 2^{\circ}$, respectively \citep{Jha2020}. 
	The properties of the impacting drop are controlled using water-glycerol mixtures with viscosities $\eta_d$ varying by almost two orders of magnitude, from $1\,\si{\milli\pascal}\si{\second}$ to $80.2\,\si{\milli\pascal}\si{\second}$. Surface tension is either $72\,\si{\milli\newton}/\si{\meter}$ (pure water) or $61\,\si{\milli\newton}/\si{\meter}$ (glycerol), while density $\rho_d$ ranges from $1000\,\si{\kilo\gram}/\si{\cubic\meter}$ to $1220\,\si{\kilo\gram}/\si{\cubic\meter}$, as detailed in table~1 \citep{cheng2008formula, volk2018density, Jha2020}.
	We note that using liquids such as silicone oil can provide a broader range of viscosity variation when paired with a superamphiphobic substrate \citep{deng2012candle}. Additionally, employing drops of smaller radii facilitates the exploration of higher Ohnesorge numbers ($Oh$, see~\eqref{Eq:OhDefinition}). 
	The drop diameter $D_0$ is controlled between $2.05\,\si{\milli\meter}$ and $2.76\,\si{\milli\meter}$ by pushing it through a calibrated needle (see appendix~\ref{app:error} for details). 
	Consequently, we calculate $Oh$ using the properties in table~\ref{tab:table00}.
	The Weber number ($We$, see~\eqref{Eq:WeDefinition}) is set using the impact velocity $V_0$ varying between $0.38\,\si{\meter}/\si{\second}$ and $2.96\,\si{\meter}/\si{\second}$ by changing the release height of the drops above the substrate. 
	All experiments are conducted at ambient pressure and temperature. The impact force is directly measured using a high-precision piezoelectric force transducer (Kistler 9215A) with a resolution of $0.5\,\si{\milli\newton}$. During these measurements, the high-frequency vibrations induced by the measurement system and the surrounding noise are spectrally removed using a low pass filter with a cut-off frequency of $5\,\si{\kilo\hertz}$, following the procedure in \citet{Li2014, Zhang2017, Gordillo2018, Mitchell2019}. 
	The experiment also employs a high-speed camera (Photron Fastcam Nova S12) synchronized at $10,000$ fps with a shutter speed $1/20,000\,\si{\second}$. Throughout the manuscript, the error bars are of statistical nature (one standard deviation) and originate from repeated trials. They are visible if they are larger than the marker size. We refer the readers to the supplementary material of \citet{zhang2022impact}  and appendix~\ref{app:error} for further details of the experimental setup and error characterization of the dimensionless control parameters.
	
	\subsection{Numerical framework}
	
	In the direct numerical simulations (DNS) employed for this study, the continuity and the momentum equations take the form
	
	\begin{align}\label{eqn:continuity}
		\boldsymbol{\nabla\cdot v} = 0 
	\end{align}
	
	\noindent and
	
	\begin{align}
		\frac{\partial \boldsymbol{v}}{\partial t} + \boldsymbol{\nabla \cdot}\left(\boldsymbol{vv}\right) = \frac{1}{\rho}\left(-\boldsymbol{\nabla} p + \boldsymbol{\nabla\cdot}\left(2\eta\boldsymbol{\mathcal{D}}\right)  + \boldsymbol{f}_\gamma\right) + \boldsymbol{g},
	\end{align}
	
	\noindent respectively. Here, $\boldsymbol{v}$ is the velocity field, $t$ is time, $p$ is pressure, and $\boldsymbol{g}$ is acceleration due to gravity. We use the free software program \textit{Basilisk C} that employs the well-balanced geometric volume of fluid (VoF) method \citep{popinet2009accurate, popinet2018numerical}. The VoF tracer $\Psi$ delineates the interface between the drop (subscript \textit{d}, $\psi = 1$)  and air (subscript \textit{a}, $\psi = 0$), introducing a singular force $\boldsymbol{f}_\gamma \approx \gamma\kappa\boldsymbol{\nabla}\Psi$ \citep[$\kappa$ denotes interfacial curvature, see][]{brackbill1992continuum} to respect the dynamic boundary condition at the interface. This VoF tracer sets the material properties such that density $\rho$ and viscosity $\eta$ are given by
	
	\begin{align}
		\rho = \rho_a + \left(\rho_d - \rho_a\right)\Psi
	\end{align}
	
	\noindent and
	
	\begin{align}
		\eta = \eta_a + \left(\eta_d - \eta_a\right)\Psi,
	\end{align}
	
	\noindent respectively. This VoF field is advected with the flow, following the equation
	
	\begin{align}\label{eqn:VoF}
		\frac{\partial \Psi}{\partial t} + \boldsymbol{\nabla \cdot}\left(\boldsymbol{v}\Psi\right) = 0.
	\end{align}

	\noindent Lastly, we calculate the normal reaction force $\boldsymbol{F}(t)$ by integrating the pressure field $p$ at the substrate,
	
	\begin{align}\label{Eqn:ReactionForce}
		\boldsymbol{F}(t) = \left(\int_\mathcal{A} \left(p-p_0\right)d\mathcal{A}\right)\hat{\boldsymbol{z}},
	\end{align}
	
	\noindent where, $p_0$, $d\mathcal{A}$, and $\hat{\boldsymbol{z}}$ are the ambient pressure, substrate area element, and the unit vector normal to the substrate, respectively. 
	
	We leverage the axial symmetry of the drop impact (figure~\ref{fig:schematic}a). This axial symmetry breaks at large $We$ ($\ge 100$ for water drops and even larger Weber number for more viscous drops), owing to destabilization by the surrounding gas after splashing \citep{xu2005drop, Eggers2010, Driscoll2011, riboux2014experiments, Josserand2016, zhang2022impact}. 
	To solve the governing equations~\eqref{eqn:continuity}-\eqref{eqn:VoF}, the velocity field $\boldsymbol{v}$ and time $t$ are normalized by the inertio-capillary scales, $V_\gamma = \sqrt{\gamma/\rho_dD_0}$ and $\tau_\gamma = \sqrt{\rho_dD_0^3/\gamma}$, respectively. Furthermore, the pressure is normalized using the capillary pressure scale $p_\gamma = \gamma/D_0$. In such a conceptualization, $Oh$ and $We$ described in \S\ref{sec:intro} uniquely determine the system. 
	The Ohnesorge number based on air viscosity $Oh_a = \left(\eta_a/\eta_d\right)Oh$ and air-drop density ratio $\rho_a/\rho_d$ are fixed at $10^{-5}$ and $10^{-3}$, respectively to minimize the influence of the surrounding medium on the impact forces. 
	Lastly, we keep the Bond number $Bo$ (see~\eqref{Eq:BoDefinition}) fixed at 1 throughout the manuscript. In our system, the relevance of gravity is characterized by the dimensionless Froude number $Fr = V_0^2/(gD_0) = We/Bo$ which compares inertia with gravity. Throughout this manuscript, $Fr > 1$ and gravity's role is sub-dominant compared to inertia (for detailed discussion, see appendix~\ref{app:gravity}).
	The substrate is modeled as a no-slip and non-penetrable wall, whereas vanishing stress and pressure are applied at the remaining boundaries to mimic outflow conditions for the surrounding air. The domain boundaries are far enough from the drop not to influence its impact process ($\mathcal{L}_\text{max} \gg D_0$, $\mathcal{L}_\text{max} = 8R$ in the worst case). 
	At $t = 0$, in our simulations, we release a spherical drop whose south pole is $0.05D_0$ away from the substrate and is falling with a velocity $V_0$. 
	It is important to note that large experimental drops may deviate from perfect sphericity due to air drag as they fall after detaching from the needle and potential residual oscillations from detachment. These shape perturbations are more pronounced in cases with low Weber and Ohnesorge numbers. To quantify this non-sphericity, we measure the drop's aspect ratio (horizontal to vertical diameter) immediately before substrate contact. The precise pre-impact drop shape can significantly influence subsequent impact dynamics \citep{thoraval-2013-jfm, yun2017bouncing,Zhang2019}. In our experiments, we constrain our analysis to drops with aspect ratios between 0.96 and 1.05. Given this narrow range, we posit that the impact of these shape variations is negligible compared to the experimental error bars derived from repeated trials under identical nominal conditions.
	The simulations utilize adaptive mesh refinement to finely resolve the velocity, viscous dissipation, and the VoF tracer fields. A minimum grid size $\Delta = D_0/2048$ is used for this study. 
	
	To ensure a perfectly non-wetting surface, we impose a thin air layer (minimum thickness $\sim \Delta/2$) between the drop and the substrate. This air layer prevents direct contact between the liquid and solid \citep{kolinski-2014-epl, sprittles2024gas}, effectively mimicking a perfectly non-wetting surface. The presence of this air layer is crucial for capturing the dynamics of drop impact on superhydrophobic surfaces, as it allows for the formation of an air cushion that can significantly affect the spreading and rebound behavior of the drop \citep{ramirez2020lifting, sanjay_chantelot_lohse_2023}.
	While this approach does not fully resolve the microscopic dynamics within the air layer itself, such as the high-velocity gradients and viscous dissipation inside the gas film once it thins below a critical size ($\sim 10\Delta$), it has been shown to accurately capture the macroscopic behavior of drop impact in the parameter range of interest \citep{ramirez2020lifting, sanjay2023drop, alventosa2023inertio, garcia2024skating}. We refer the readers to \citet{VatsalThesis} for discussions about this \lq\lq precursor\rq\rq, air film method and to \citet{popinet-basilisk, basiliskVatsal, zhang2022impact} for details on the numerical framework.
	
	\section{Anatomy of the first impact force peak}\label{sec:FirstPeak}
	
	This section elucidates the anatomy of the first impact force peak and its relationship with the Weber $We$ and Ohnesorge $Oh$ numbers, first for the inertial limit (\S\ref{sec:FirstInertialPeak}, $Oh \ll 1$) and then for the viscous asymptote (\S\ref{sec:ViscousImpact}, $Oh \gg 1$). The results of this section are summarized in figure~\ref{fig:F1Anatomy} that shows an excellent agreement between experiments and simulations without any free parameters. 
	
	\subsection{Low Ohnesorge number impacts}\label{sec:FirstInertialPeak}
	\begin{figure}
		\centering
		\includegraphics[width=\textwidth]{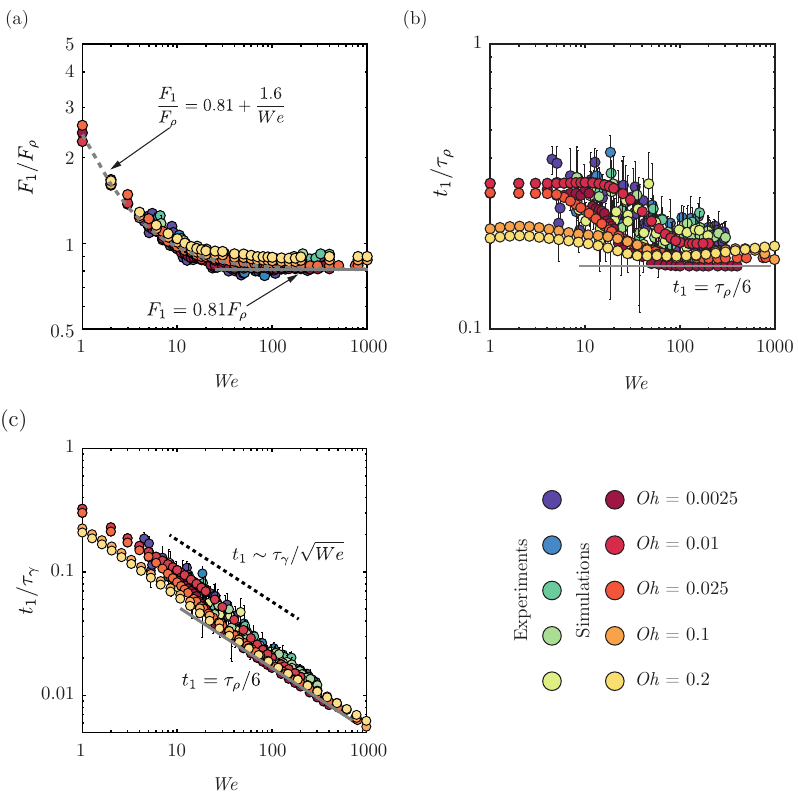}
		\caption{Anatomy of the first impact force peak amplitude at low $Oh$ in between 0.0025 and 0.2, see color legend: $We$ dependence of the (a) magnitude $F_1$ normalized by the inertial force scale $F_\rho = \rho_dV_0^2D_0^2$ and time $t_1$ to reach the first force peak amplitude normalized by (b) the inertial timescale $\tau_\rho = D_0/V_0$ and (c) the inertio-capillary time scale $\tau_\gamma = \sqrt{\rho_dD_0^3/\gamma}$.}
		\label{fig:F1Anatomy}
	\end{figure}
	\begin{figure}
		\centering
		\includegraphics[width=\textwidth]{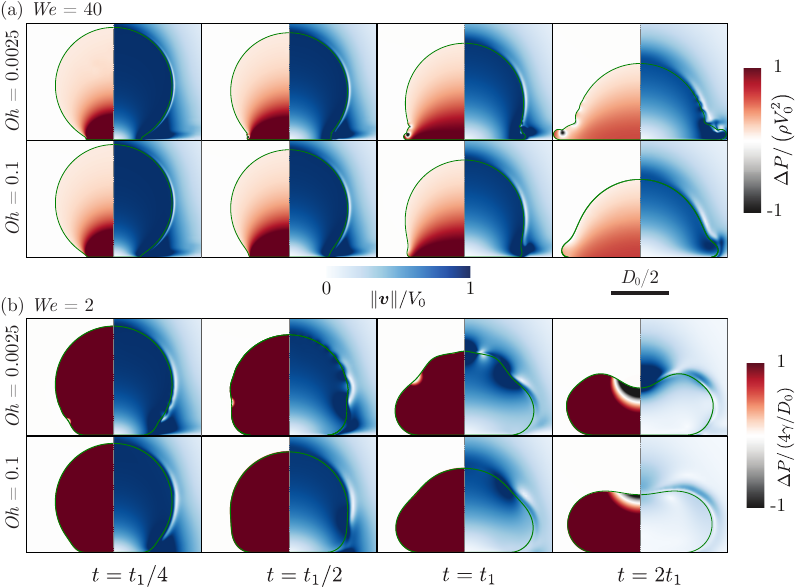}
		\caption{Direct numerical simulations snapshots illustrating the drop impact dynamics for $We$ = (a) 40 and (b) 2. The left-hand side of each numerical snapshot shows the pressure normalized by (a) the inertial pressure scale $\rho_dV_0^2$ and (b) the capillary pressure scale $\gamma/D_0$. The right-hand side shows the velocity field magnitude normalized by the impact velocity $V_0$.}
		\label{fig:F1AnatomyLowOh}
	\end{figure}
	
	For low $Oh$ and large $We$, inertial force and time scales dictate the drop impact dynamics (figures~\ref{fig:F1Anatomy} and~\ref{fig:F1AnatomyLowOh}). As the drop falls on a substrate, the part of the drop immediately in contact with the substrate stops moving, whereas the top of the drop still falls with the impact velocity (figure~\ref{fig:F1AnatomyLowOh}, from $t = t_1/4$ until $t = t_1$). Consequently, momentum conservation implies 
	
	\begin{align}
		F_1 \sim V_0\frac{dm}{dt},
	\end{align}
	
	\noindent where the mass flux $dm/dt \sim \rho_d V_0D_0^2$ \citep{Soto2014, zhang2022impact}. As a result, the first peak amplitude scales with the inertial pressure force (figure~\ref{fig:F1Anatomy}a)
	
	\begin{align}\label{eq:F1}
		F_1 \sim F_\rho,\text{ where }F_\rho =  \rho_d V_0^2D_0^2,
	\end{align}
	
	\noindent for high Weber numbers ($\Wen > 30$, $F_1 \approx 0.81F_\rho$). Furthermore, the time $t_1$ to reach $F_1$ follows 
	
	\begin{align}\label{eq:t1}
		t_1 \sim \frac{D_0}{V_0} = \tau_\rho,
	\end{align}
	
	\noindent where, $\tau_\rho$ is the inertial timescale. The relation between equations~\eqref{eq:F1} and~\eqref{eq:t1} is apparent from the momentum conservation which implies that the impulse of the first force peak is equal to the momentum of the impacting drop, i.e., $F_1t_1 \sim \rho_dV_0D_0^3 = F_\rho\tau_\rho$ (see \citealp{Gordillo2018}, \citealp{zhang2022impact}, and figures~\ref{fig:F1Anatomy}b,c). These scaling laws depend only on the inertial shock at impact and are wettability-independent \citep{Zhang2017, Gordillo2018, zhang2022impact}. For details of the scaling law, including the prefactors, we refer the readers to \citet{Philippi2016, Gordillo2018, cheng2021drop}. 
	
	Figure~\ref{fig:F1Anatomy} further illustrates that this inertial asymptote is insensitive to viscosity variations up to 100-fold as $F_1 \sim F_\rho$ and $t_1 \sim \tau_\rho$ for $0.0025 < Oh < 0.2$. However, deviations from the inertial force and time scales are apparent for $\Wen < 30$ (figure~\ref{fig:F1Anatomy}), a phenomenon also reported in earlier work \citep{Soto2014, zhang2022impact}. In these instances, inertia does not act as the sole governing force but instead complements surface tension, which dictates the pressure inside the drop ($p \sim \gamma/D_0$ throughout the drop for $We \lesssim 1$, figure~\ref{fig:F1AnatomyLowOh}b). \citet{zhang2022impact} proposed an empirical functional dependence as
	
	\begin{align}
		F_1 = \left(\alpha_1\rho_d V_0^2 + \alpha_2\frac{\gamma}{D_0}\right)D_0^2,
	\end{align}
	
	\noindent based on dimensional analysis, with $\alpha_1$ and $\alpha_2$ as free parameters which were determined to be approximately $1.6$ and $0.81$, respectively for water ($Oh = 0.0025$). These coefficients only deviate marginally in the current work despite the significant increase in $Oh$ as compared to previous works \citep{cheng2021drop, zhang2022impact}. This consistency underscores the invariance of the pressure field inside the drop to an increase in $Oh$ (close to the impact region, figure~\ref{fig:F1AnatomyLowOh}a and throughout the drop, figure~\ref{fig:F1AnatomyLowOh}b).
	
	\subsection{Large Ohnesorge number impacts}\label{sec:ViscousImpact}
	
	\begin{figure}
		\centering
		\includegraphics[width=\textwidth]{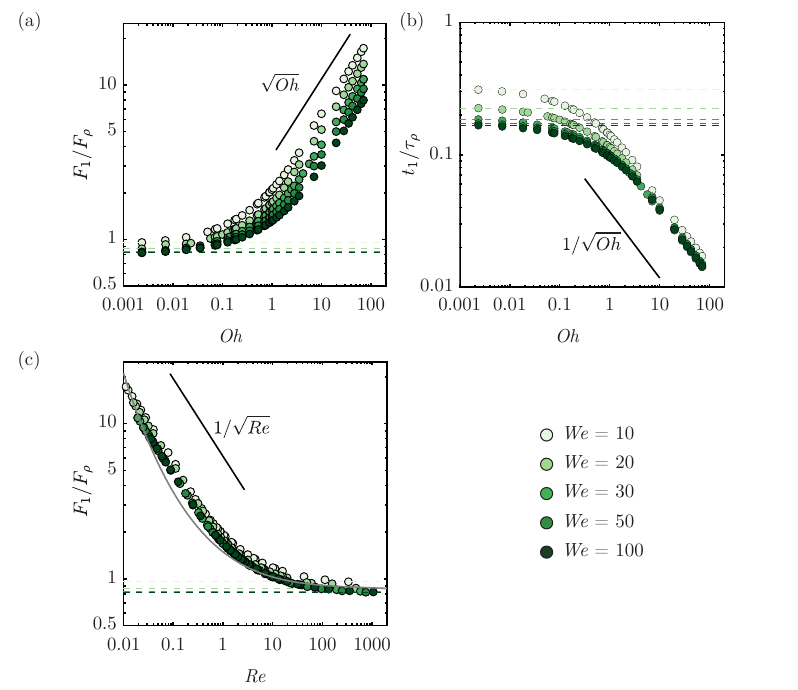}
		\caption{Anatomy of the first impact force peak amplitude for viscous impacts from our numerical simulations: the $Oh$ dependence of  (a) the magnitude $F_1$ normalized by the inertial force scale $\rho_dV_0^2D_0^2$ and (b) the time $t_1$ to reach the first force peak amplitude normalized by inertial timescale $\tau_\rho = D_0/V_0$. (c) The $Re$ dependence of the magnitude $F_1$ normalized by the inertial force scale $\rho_dV_0^2D_0^2$ as compared to the (implicit) theoretical calculation of \citet{Gordillo2018}. The black line corresponds to the scaling relationship described in \S\ref{sec:ViscousImpact}. The Weber number is color-coded.}
		\label{fig:F1Anatomy_2}
	\end{figure}
	
	\begin{figure}
		\centering
		\includegraphics[width=\textwidth]{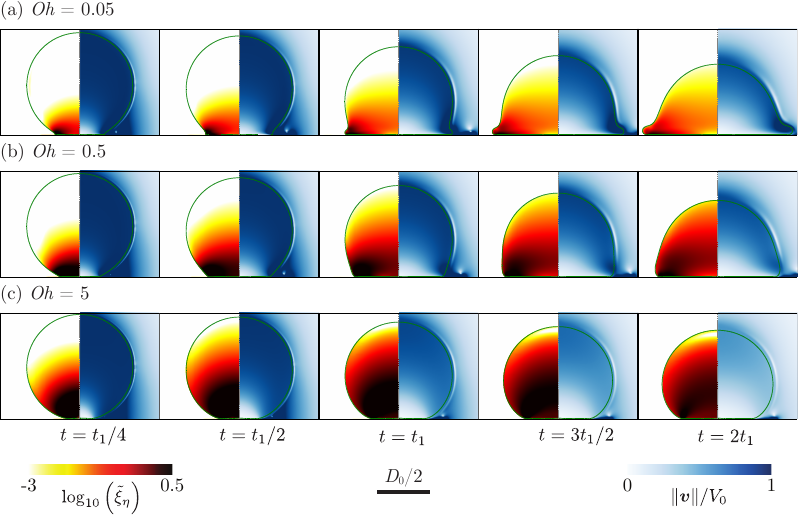}
		\caption{Direct numerical simulations snapshots illustrating the drop impact dynamics for $We = 100$ and $Oh$ = (a) 0.05, (b) 0.5, and (c) 5. The left-hand side of each numerical snapshot shows the viscous dissipation function $\xi_\eta$ normalized by the inertial scale $\rho_dV_0^3/D_0$. The right-hand side shows the velocity field magnitude normalized by the impact velocity $V_0$.}
		\label{fig:QualitativeF1}
	\end{figure}
	
	Figure~\ref{fig:F1Anatomy_2} reaffirms the findings of \S\ref{sec:FirstInertialPeak} for low $Oh$ that the first impact peak amplitude $F_1$ and the time to reach this peak amplitude $t_1$ scale with $F_\rho$ and $\tau_\rho$, respectively. As the Ohnesorge number increases further, the first impact force peak amplitude normalized with $F_\rho$ begins to increase, indicating a transition around $Oh \approx 0.1$, where viscosity starts to play a significant role. At large $Oh$, we observe the scaling relationship (figure~\ref{fig:F1Anatomy_2}a)
	
	\begin{align}
		F_1 \sim F_\rho\sqrt{Oh}.
	\end{align}
	
	\noindent The drop's momentum is still $\rho_dV_0D_0^3$ which must be balanced by the impulse from the substrate, $F_1t_1$ (see \S\ref{sec:FirstInertialPeak}, \citealp{Gordillo2018}, \citealp{zhang2022impact}, and figure~\ref{fig:F1Anatomy_2}b). Consequently, the time $t_1$ follows
	
	\begin{align}
		t_1 \sim \frac{\tau_\rho}{\sqrt{Oh}}.
	\end{align}
	
	Figure~\ref{fig:F1Anatomy_2} further shows that these scaling laws are weakly dependent on the Weber number, as viscous dissipation consumes the entire initial kinetic energy of the impacting drop (figure~\ref{fig:QualitativeF1}). Once again, we stress that using the water-glycerol mixtures limits the range of $Oh$ that we can probe experimentally. 
	We further note that the first peak is robust and does not depend on the wettability of the substrate. Consequently, to compare with the existing data such as those in \citet{cheng2021drop} with different liquids to cover a wider range of liquid viscosities and to account for the apparent $We$-dependence, we plot $F_1$ compensated with $F_\rho$ against the impact Reynolds number $Re \equiv \sqrt{We}/Oh = V_0D_0/\nu_d$. 
	For the low $Re$ regime, such a plot allows us to describe the $We$ dependence on the prefactor more effectively, as illustrated in figure~\ref{fig:F1Anatomy_2}(c). However, it is important to note that some scatter is still observed at high $Re$ values, which can be attributed to the $We$ dependence of the impact force peak amplitude. This lack of a pure scaling behavior demonstrates how the interplay between kinetic energy and viscous dissipation within the drop dictates the functional dependence of the maximum impact force on $Oh$.
	
	To systematically elucidate these scaling behaviors in the limit of small $Re$, we need to find the typical scales for the rate of change of kinetic energy and that of the rate of viscous dissipation for the drop impact system. First, we can readily define an average rate of viscous dissipation per unit mass as
	
	\begin{align}
		\bar{\varepsilon} \sim \frac{1}{\tau_\rho}\frac{1}{D_0^3}\int_0^{\tau_\rho}\int_\Omega\nu_d\left(\boldsymbol{\mathcal{D}:\mathcal{D}}\right)d\Omega dt,
	\end{align}
	
	\noindent where $\nu_d$ is the kinematic viscosity of the drop and $d\Omega$ is the volume element where dissipation occurs. Notice that $\bar{\varepsilon}$ has the dimensions of $V_0^3/D_0$, i.e., length squared over time cubed or velocity squared over time, as it should be for dissipation rate of energy per unit mass. We can estimate $\Omega = D_{\text{foot}}^2l_\nu$ (figure~\ref{fig:QualitativeF1}), where $D_{\text{foot}}$ is the drop's foot diameter in contact with the substrate and $l_\nu$ is the viscous boundary layer thickness. This boundary layer marks the region of strong velocity gradients ($\sim V_0/l_\nu$) analogous to the \citet{mirels1955laminar} shockwave-induced boundary layer. For details, we refer the authors to \citet{schlichting2016boundary, Schroll2010, Philippi2016}. Consequently, the viscous dissipation rate scales as
	
	\begin{align}\label{eq:dissipationScale}
		\bar{\varepsilon} \sim \frac{1}{\tau_\rho D_0^3}\int_0^{\tau_\rho}\nu_d \left(\frac{V_0}{l_\nu}\right)^2 D_{\text{foot}}^2l_\nu dt.
	\end{align}
	
	\noindent To calculate $D_{\text{foot}}$, we assume that the drop maintains a spherical cap shape throughout the impact (figure~\ref{fig:QualitativeF1}). To calculate the distance the drop would have traveled if there were no substrate, we use the relation $d \sim V_0t$. Simple geometric arguments allow us to determine the relation between the foot diameter and this distance, $D_{\text{foot}} \sim \sqrt{D_0d}$ \citep{lesser1981analytic, mandre2009precursors,  zheng2021air, bilotto2023fluid, bertin2023similarity}. Interestingly, this scaling behavior is similar to the inertial limit \citep{wagner1932stoss, Bouwhuis2012, Philippi2016, gordillo2019theory} as discussed by \citet{langley2017impact, bilotto2023fluid}. Furthermore, the viscous boundary layer $l_\nu$ can be approximated using $\sqrt{\nu_d t}$ \citep{mirels1955laminar, Eggers2010, Philippi2016}. Filling these in \eqref{eq:dissipationScale}, we get
	
	\begin{align}	
		\bar{\varepsilon} \sim \frac{1}{\tau_\rho D_0^2}\int_0^{\tau_\rho}\sqrt{\nu_d} V_0^3 \sqrt{t} dt,
	\end{align}
	
	\noindent which on integration gives 
	
	\begin{align}\label{eq:eps0Final}
		\bar{\varepsilon} \sim \sqrt{\nu_d \tau_\rho}V_0^3/D_0^2, 
	\end{align}
	
	\noindent where $\tau_\rho$ is the inertial time scale. Here, we assume that for highly viscous drops, all energy is dissipated within a fraction of $\tau_\rho$. Filling in \eqref{eq:eps0Final} and normalizing $\bar{\varepsilon}$ with the inertial scales $V_0^3/D_0$,
	
	\begin{align}\label{eq:DissipationScale}
		\frac{\bar{\varepsilon}}{V_0^3/D_0} \sim \sqrt{\frac{\nu_d\tau_\rho}{D_0^2}} = \frac{1}{\sqrt{Re}} = \left(\frac{Oh}{\sqrt{We}}\right)^{1/2}.
	\end{align}
	
	\noindent Next, the kinetic energy of the falling drop is given by
	
	\begin{align}\label{eq:KEdissipation}
		\dot{K}(t) \equiv \frac{dK(t)}{dt} \sim \rho_dD_0^3\bar{\varepsilon},\quad\text{where } K(t) = \frac{1}{2}m\left(V(t)\right)^2,
	\end{align}
	
	\noindent and $V(t)$ is the drop's center of mass velocity. The left-hand side of \eqref{eq:KEdissipation} can be written as
	
	\begin{align}\label{eq:KE-powerTime}
		\dot{K}(t) = mV(t)\frac{dV(t)}{dt} = F(t)V(t).
	\end{align}
	
	\noindent In equation~\eqref{eq:KE-powerTime}, $F(t)$ and $V(t)$ scale with the first impact force peak amplitude $F_1$ and the impact velocity $V_0$, respectively, giving the typical scale of the rate of change of kinetic energy as
	
	\begin{align}\label{eq:KE-power}
		\dot{K}^* \sim F_1V_0.
	\end{align}
	
	\noindent We stress that \eqref{eq:KE-power} states that the rate of change of kinetic energy is equal to the power of the normal reaction force, an observation already made by \citet{wagner1932stoss} and \citet{Philippi2016} in the context of impact problems. Lastly, at large $Oh$, viscous dissipation enervates kinetic energy completely giving (figure~\ref{fig:QualitativeF1}c, also see:  \citet{Philippi2016} and \citet{ Wildeman2016}),
	
	\begin{align}\label{eq:force-Dissipation}
		\dot{K}^* \sim F_1V_0 \sim \rho_dD_0^3\bar{\varepsilon}
	\end{align}
	
	\noindent Additionally, we use the inertial scales to non-dimensionalize  \eqref{eq:force-Dissipation} and fill in \eqref{eq:DissipationScale}, giving
	
	\begin{align}
		\label{Eq:viscIncrease}
		\frac{F}{F_\rho} \sim \frac{\bar{\varepsilon}}{V_0^3/D_0} \sim \frac{1}{\sqrt{Re}} = \left(\frac{Oh}{\sqrt{We}}\right)^{1/2}
	\end{align}
	
	\noindent and using $F_1t_1 \sim \rho_dV_0D_0^3 = F_\rho\tau_\rho$,
	
	\begin{align}
		\label{Eq:viscTime}
		\frac{t_1}{\tau_\rho} \sim \left(\frac{\sqrt{We}}{Oh}\right)^{1/2}.
	\end{align}
	
	In summary, we use energy and momentum invariance to elucidate the parameter dependencies of the impact force as illustrated in figure~\ref{fig:F1Anatomy_2}. The scaling arguments capture the dominant force balance during the impact process, considering the relative importance of inertial, capillary, and viscous forces. As the dimensionless viscosity of impacting drops increases, the lack of surface deformation increases the normal reaction force \eqref{Eq:viscIncrease}. Further, the invariance of incoming drop momentum implies that this increase in normal reaction force occurs on a shorter timescale \eqref{Eq:viscTime}.
	
	\section{Anatomy of the second impact force peak}\label{sec:SecondPeak}
	
	\begin{figure}
		\centering
		\includegraphics[width=\textwidth]{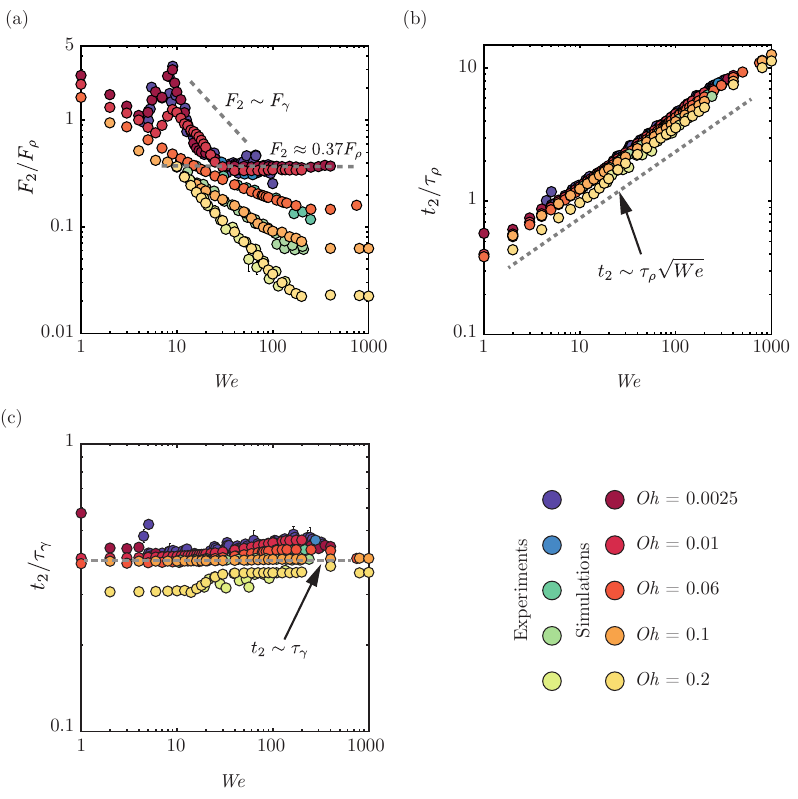}
		\caption{Anatomy of the second impact force peak amplitude: $We$ dependence of the (a) magnitude $F_2$ normalized by the inertial force scale $F_\rho = \rho_dV_0^2D_0^2$ and time $t_2$ to reach the second force peak amplitude normalized by (b) the inertio-capillary time scale $\tau_\gamma = \sqrt{\rho_dD_0^3/\gamma}$ and (c) inertial timescale $\tau_\rho = D_0/V_0$.}
		\label{fig:F2Anatomy}
	\end{figure}
	
	\begin{figure}
		\centering
		\includegraphics[width=\textwidth]{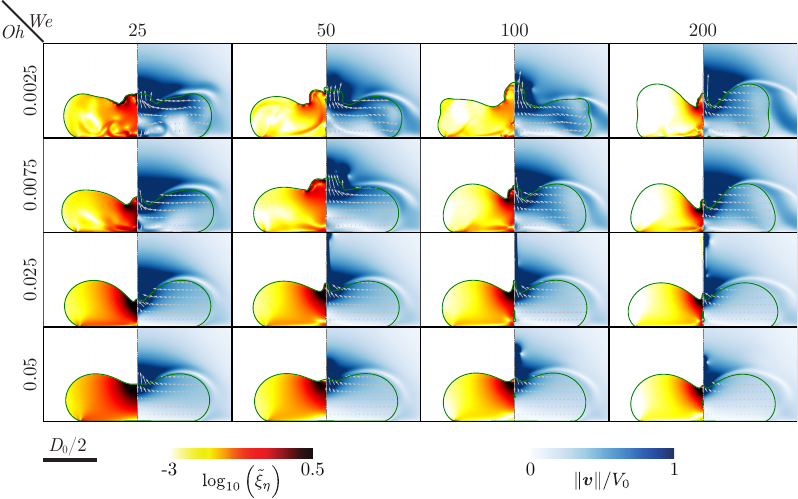}
		\caption{Direct numerical simulations snapshots illustrating the influence of $We$ and $Oh$ on the inception of the Worthington jet. All these snapshots are taken at the instant when the second peak appears in the temporal evolution of the normal reaction force ($t = t_2$). The left-hand side of each numerical snapshot shows the viscous dissipation function $\xi_\eta$ normalized by the inertial scale $\rho_dV_0^3/D_0$. The right-hand side shows the velocity field magnitude normalized by the impact velocity $V_0$. The gray velocity vectors are plotted in the center of mass reference frame of the drop to clearly elucidate the internal flow.}
		\label{fig:F2Phenomenology}
	\end{figure}
	
	\begin{figure}
		\centering
		\includegraphics[width=\textwidth]{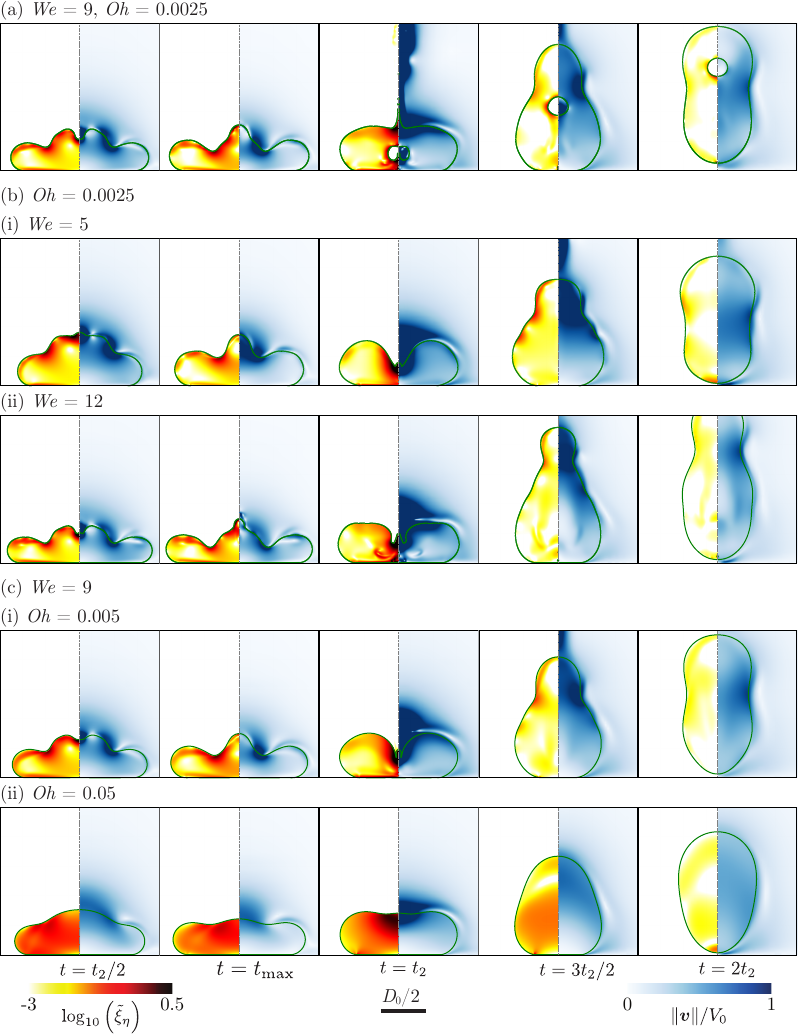}
		\caption{Direct numerical simulations snapshots illustrating the influence of $We$ and $Oh$ on the singular Worthington jet. (a) ($We, Oh$) = ($9, 0.0025$), (b) $Oh = 0.0025$ with $We$ = (i) $5$ and (ii) $12$, and (c) $We = 9$ with $Oh$ = (i) $0.005$ and (ii) $Oh = 0.05$. The left-hand side of each numerical snapshot shows the viscous dissipation function $\xi_\eta$ normalized by inertial scale $\rho_dV_0^3/D_0$. The right-hand side shows the velocity field magnitude normalized by the impact velocity $V_0$.}
		\label{fig:KillSingularity}
	\end{figure}
	
	This section delves into the anatomy of the second impact force peak amplitude $F_2$ as a function of the Weber $We$ and Ohnesorge $Oh$ numbers, summarized in figure~\ref{fig:F2Anatomy}. We once again note the remarkable agreement between  experiments and numerical simulations in this figure. 
	
	Similar to the mechanism leading to the formation of the first peak (\S\ref{sec:FirstPeak}), also the mechanism for the formation of this second peak is momentum conservation. As the drop takes off the surface, it applies a force on the substrate. As noted in \S\ref{sec:intro} and \citet{zhang2022impact}, this force also coincides with the formation of a Worthington jet (figure~\ref{fig:summary}iv-vi). The time $t_2$ at which the second peak is observed scales with the inertio-capillary timescale and is insensitive to $We$ and $Oh$ (figure~\ref{fig:F2Anatomy}b,c). Once again, we invoke the analogy between drop oscillation and drop impact to explain this behavior  \citep{Richard2002, chevy2012liquid}. At the time instant $t_2 \approx 0.44\tau_\gamma$, the drop's internal motion undergoes a transition from a predominantly radial flow to a vertical one due to the formation of the Worthington jet \citep{chantelot2018rebonds, zhang2022impact}. Figure~\ref{fig:F2Phenomenology} exemplifies this jet in the $Oh$-$We$ parameter space, which is intricately related to the second peak in the drop impact force. For low $Oh$ and large $We$, the drop retraction follows a modified Taylor-Culick dynamics \citep{bartolo2005retraction, Eggers2010, sanjay2022TC}. As $We$ is increased, the jet gets thinner but faster, maintaining a constant momentum flux $\rho_dV_j^2d_j^2$, where $V_j$ and $d_j$ are the jet's velocity and diameter, respectively \citep[figure~\ref{fig:F2Phenomenology},][]{zhang2022impact}. This invariance leads to the observed scaling $F_2 \sim F_\rho$ in this regime ($F_2 \approx 0.37F_\rho$ for $We \ge 30, Oh \le 0.01$).
	
	Furthermore, the low $We$ and $Oh$ regime relies entirely on capillary pressure (figure~\ref{fig:F1AnatomyLowOh}). Subsequently, $F_2 \sim F_\gamma = \gamma D_0$ for $Oh < 0.01$ and $We < 30$ (figure~\ref{fig:F2Anatomy}). This flow focusing (figure~\ref{fig:F2Phenomenology}) is most efficient for $We = 9$ (figure~\ref{fig:KillSingularity}a, $t_2/2 < t < t_2$, \citealp{renardy2003pyramidal, Bartolo2006Singular}) where the capillary resonance leads to a thin-fast jet, accompanied by a bubble entrainment, reminiscent of the hydrodynamic singularity (figure~\ref{fig:KillSingularity}, \citealp{zhang2022impact, sanjay_lohse_jalaal_2021}). The characteristic feature of this converging flow is a higher magnitude of $F_2$ compared to $F_1$ (figure~\ref{fig:F2Anatomy}). 
	
	However, this singular jet regime is very narrow in the $Oh$-$We$ phase space. Figure~\ref{fig:KillSingularity}b shows two cases for water drops ($Oh = 0.0025$) at different $We$ ($5$ and $12$ for figures~\ref{fig:KillSingularity}b-i and b-ii, respectively). Bubble entrainment does not occur in either of these cases. Consequently, the maximum force amplitude diminishes for these two cases (figure~\ref{fig:F2Anatomy}). Nonetheless, these cases are still associated with high local viscous dissipation near the axis of symmetry owing to the singular nature of the flow. Another mechanism to inhibit this singular Worthington jet is viscous dissipation in the bulk. As the Ohnesorge number increases, this singular jet formation disappears ($Oh = 0.005$, figure~\ref{fig:KillSingularity}c-i), significantly reducing the second peak of the impact force. For even higher viscosities, the drop no longer exhibits the sharp, focused jet formation seen at lower viscosities, and the second peak in the force is notably diminished ($Oh = 0.05$, figure~\ref{fig:KillSingularity}c-ii).
	
	Lastly,  as $Oh$ increases, bulk dissipation becomes dominant (apparent from increasing $Oh$ at fixed $We$ in figure~\ref{fig:F2Phenomenology}) and can entirely inhibit drop bouncing. Recently, \citet{Jha2020, sanjay_chantelot_lohse_2023} showed that there exists a critical $Oh$, two orders of magnitude higher than that of a $2\,\si{\milli\meter}$ diameter water drop, beyond which drops do not bounce either, irrespective of their impact velocity. Consequently, the second peak in the impact force diminishes for larger $Oh$, which explains the monotonic decrease of the amplitude $F_2$ observed in figure~\ref{fig:F2Anatomy} for $We > 30, Oh > 0.01$.
	
	\section{Conclusion and outlook}\label{sec:Conclusion}
	
	\begin{figure}
		\centering
		\includegraphics[width=\textwidth]{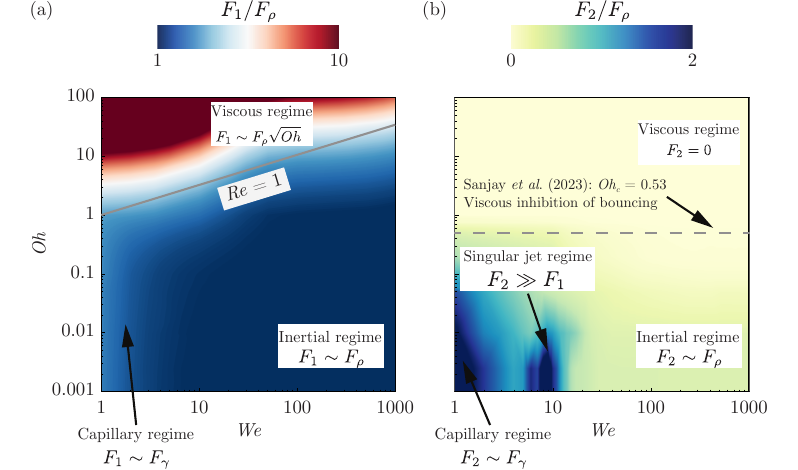}
		\caption{Regime map in terms of the drop Ohnesorge number $Oh$ and the impact Weber number $We$ to summarize the two peaks in the impact force by showing the different regimes described in this work based on (a) the first peak in the impact force peak amplitude $F_1$ and (b) the second peak in the impact force peak amplitude $F_2$. Both peaks are normalized by the inertial force scale $F_\rho = \rho_dV_0^2D_0^2$. These regime maps are constructed using $\sim 1500$ simulations in the range $0.001 \leq Oh \leq 100$ and $1 \leq We \leq 1000$. The gray solid line in (a) and dashed line in (b) mark the inertial--viscous transition ($Re = 1$) and the bouncing--no-bouncing transition \citep[$Oh_c = 0.53$ for $Bo = 1$, see][]{sanjay_chantelot_lohse_2023}, respectively.}
		\label{fig:RegimeMaps}
	\end{figure}
	
	In this work, we study the forces and dissipation encountered during the drop impact process by employing experiments, numerical simulations, and theoretical scaling laws. We vary the two dimensionless control parameters--the Weber ($We$: dimensionless impact kinetic energy) and the Ohnesorge number ($Oh$: dimensionless viscosity) independently to elucidate the intricate interplay between inertia, viscosity, and surface tension in governing the forces exerted by a liquid drop upon impact on a non-wetting substrate.
	
	For the first impact force peak amplitude $F_1$, owing to the momentum balance after the inertial shock at impact, figure~\ref{fig:RegimeMaps}(a) summarizes the different regimes in the $Oh$-$We$ phase space. For low $Oh$, inertial forces predominantly dictate the impact dynamics, such that $F_1$ scales with the inertial force $F_\rho$ \citep{Philippi2016, Gordillo2018, Mitchell2019, cheng2021drop, zhang2022impact} and is insensitive to viscosity variations up to 100-fold. As $Oh$ increases, the viscosity becomes significant, leading to a new scaling law: $F_1 \sim F_\rho\sqrt{Oh}$. The paper unravels this viscous scaling behavior by accounting for the loss of initial kinetic energy owing to viscous dissipation inside the drop. Lastly, at low $We$, the capillary pressure inside the drop leads to the scaling $F_1 \sim F_\gamma$ \citep{molavcek2012quasi, chevy2012liquid}. 
	
	The normal reaction force described in this work is responsible for deforming the drop as it spreads onto the substrate, where it stops thanks to surface tension. If the substrate is non-wetting, it retracts to minimize the surface energy and finally takes off \citep{Richard2000}. In this case, the momentum conservation leads to the formation of a Worthington jet and a second peak in the normal reaction force, as summarized in figure~\ref{fig:RegimeMaps}(b). For low $Oh$ and high $We$, the second force peak amplitude scales with the inertial force ($F_\rho$), following a modified Taylor-Culick dynamics \citep{Eggers2010}. In contrast, capillary forces dominate at low $We$ and low $Oh$, leading to a force amplitude scaling of $F_2 \sim F_\gamma$. We also identify a narrow regime in the $Oh$-$We$ phase space where a singular Worthington jet forms, significantly increasing $F_2$ \citep{Bartolo2006Singular, zhang2022impact}, localized in the parameter space for $We \approx 9$ and $Oh < 0.01$. As $Oh$ increases, bulk viscous dissipation counteracts this jet formation, diminishing the second peak and ultimately inhibiting drop bouncing. 
	
	Our findings have far-reaching implications, not only enriching the fundamental understanding of fluid dynamics of drop impact but also informing practical applications in diverse fields such as inkjet printing, public health, agriculture, and material science where the entire range of $Oh$-$We$ phase space is relevant (figures~\ref{fig:schematic}b and~\ref{fig:RegimeMaps}). While this has identified new scaling laws, it also opens avenues for future research. For instance, it would be interesting to use the energy accounting approach to unify the scaling laws for the maximum spreading diameter for arbitrary $Oh$ \citep{laan2014maximum, Wildeman2016}. Although, the implicit theoretical model summarized in \citet{cheng2021drop} describes most of data in figure~\ref{fig:F1Anatomy_2}, we stress the importance of having a predictive model to determine $F_1$ for given $We$ and $Oh$ \citep{sanjay2024PRL}. The $We$ influence on the impact force also warrants further exploration, especially in the regime $We \ll 1$ for arbitrary $Oh$ \citep{chevy2012liquid, molavcek2012quasi} and drop impact on compliant surfaces \citep{alventosa_cimpeanu_harris_2023, ma2023scaling}. Another potential extension of this work is to non-Newtonian fluids \citep{martouzet2021dynamic, aguero2022impact, bertin2023similarity, jin2023marbles}.\\[5mm] 
	
	\noindent{\bf  Code availability\bf{.}} The codes used in the present article, and the parameters and data to reproduce figures~\ref{fig:F1Anatomy} and~\ref{fig:F2Anatomy} are permanently available at \citet{basiliskVatsal}.\\
	
	\noindent{\bf Acknowledgements\bf{.}}  We would like to thank Vincent Bertin for comments on the manuscript. We would also like to thank Andrea Prosperetti, Pierre Chantelot, Devaraj van der Meer, and Uddalok Sen for illuminating discussions.\\
	
	\noindent{\bf Funding\bf{.}} We acknowledge the funding by the ERC Advanced Grant No. 740479-DDD and NWO-Canon grant FIP-II. B.Z and C.L are grateful for the support from National Natural Science Foundation of China (Grant No. 12172189, 11921002, 11902179). This work was carried out on the national e-infrastructure of SURFsara, a subsidiary of SURF cooperation, the collaborative ICT organization for Dutch education and research. This work was sponsored by NWO - Domain Science for the use of supercomputer facilities.\\
	
	\noindent{\bf Declaration of Interests\bf{.}} The authors report no conflict of interest.\\
	
	\noindent{\bf  Authors' ORCID\bf{.}}  \\	
	V. Sanjay \href{https://orcid.org/0000-0002-4293-6099}{https://orcid.org/0000-0002-4293-6099};\\
	B. Zhang \href{https://orcid.org/0000-0001-8550-2584}{https://orcid.org/0000-0001-8550-2584};\\	
	C. Lv \href{https://orcid.org/0000-0001-8016-6462}{https://orcid.org/0000-0001-8016-6462};\\
	D. Lohse \href{https://orcid.org/0000-0003-4138-2255}{https://orcid.org/0000-0003-4138-2255}. \\
	
	\appendix
	\renewcommand{\thefigure}{\Alph{section}\,\arabic{figure}}
	\setcounter{figure}{0}

	\section{Note on the error characterization for the control parameters}
	\label{app:error}
	
	This appendix outlines the methodology for characterizing experimental errors in quantification of the drop's size and impact velocities which is crucial for accurate calculation of the dimensionless control parameters $We$ and $Oh$.
	The drop diameter determination involves multiple steps. First, we measure the total mass ($M_{100}$) of 100 drops using an electric balance. From this mass, using the liquid density and assuming spherical shape, we calculated the drop diameter ($D_0$). 
	We repeated this process five times, yielding $D_{0,1}$ through $D_{0,5}$. The average of these measurements provided the final drop diameter ($D_0$) and its standard error.
	For impact velocity determination, we extracted data from experimental high-speed imagery. By tracking the drop center's position in successive frames prior to substrate contact, and knowing the frame rate, we calculated the impact velocity. We repeated this process for five trials, obtaining $V_{0,1}$ through $V_{0,5}$. The average of these values gave the final impact velocity ($V_0$) and its standard error.
	
	The standard errors for drop diameters do not exceed $0.13\,\si{\milli\meter}$. For instance, drops with Ohnesorge numbers of $0.0025$, $0.06$, and $0.2$ have diameters of $2.05 \pm 0.13\,\si{\milli\meter}$, $2.52 \pm 0.11\,\si{\milli\meter}$, and $2.54 \pm 0.09\,\si{\milli\meter}$, respectively. 
	The standard errors for impact velocities did not exceed $0.02\,\si{\meter}/\si{\second}$. For the same $Oh$ values, the impact velocities were $1.2 \pm 0.002\,\si{\meter}/\si{\second}$, $0.97 \pm 0.01\,\si{\meter}/\si{\second}$, and $0.96 \pm 0.01\,\si{\meter}/\si{\second}$, respectively.
	The combined errors in $D_0$ and $V_0$ resulted in approximately $\pm 7\%$ error in Weber number $We$ and $\pm 3\%$ error in Ohnesorge number $Oh$. Consequently, the horizontal error bars, which relate to errors in the control parameters, are smaller than the symbol sizes in our figures.

	\begin{figure}
		\centering
		\includegraphics[width=\textwidth]{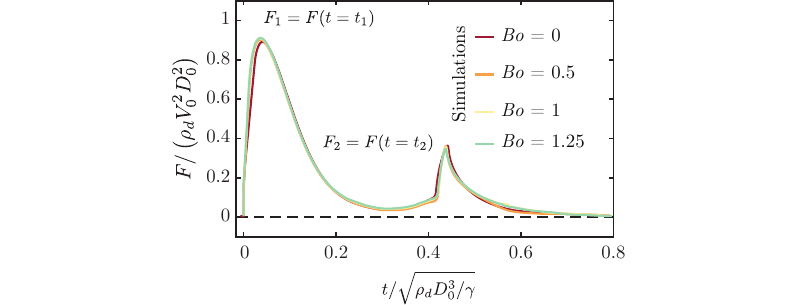}
		\caption{Comparison of the drop impact force $F(t)$ obtained from simulations for the four different Bond numbers $Bo = 0, 0.5, 1, 1.25$. Here, $Oh = 0.06$ and $We = 40$. Both force peaks $F_1$ and $F_2$ as well as time to reach these peaks $t_1$ and $t_2$ are invariant to variation in $Bo$.}
		\label{fig:AppGravity}
	\end{figure}
	
	\section{Role of gravity on drop impact forces}
	\label{app:gravity}
	
	Following table~\ref{tab:table00} and considering the variation in impacting drop diameter (appendix~\ref{app:error}), the Bond number (equation~\eqref{Eq:BoDefinition}) in our experiments ranges from $0.5$ to $1.25$, introducing an additional dimensionless control parameter alongside $We$ and $Oh$. 
	Gravity typically plays a negligible role in these impact processes \citep{sanjay_chantelot_lohse_2023,sanjay2024PRL}. We undertook a sensitivity test varying the Bond number from $0$ to $1.25$ in our simulations. 
	Figure~\ref{fig:AppGravity} confirms the leading-order Bond invariance of the results as the impact force profiles, including both force peaks $F_1$ and $F_2$ and their corresponding times $t_1$ and $t_2$, remain invariant to these Bond number variations.
	Notably, while gravity does play a role in drop impact dynamics, particularly for longer time scales and in determining the critical Ohnesorge number $Oh_c$ for bouncing inhibition (see figure~\ref{fig:RegimeMaps}b and \citet{sanjay_chantelot_lohse_2023}), its effect on the initial impact force peaks is minimal for the parameter range studied here (large Froude numbers, $Fr > 1$). This Bond number invariance allows us to focus on the more dominant effects of Weber and Ohnesorge numbers.
	Consequently, we selected the representative value of $Bo = 1$, corresponding to a diameter of $0.00254\,\si{\milli\meter}$, density $1000\,\si{\kilogram/\meter^3}$, gravitational acceleration $10\,\si{\meter/\second^2}$, and surface tension $0.061\,\si{\newton/\meter}$.

	\bibliographystyle{jfm}
	\bibliography{ViscousDropImpact_v2}
	
\end{document}